\documentclass[11pt]{article}
\usepackage{graphicx}
\usepackage{longtable,lscape}
\usepackage{amsthm}
\usepackage{amsfonts}
\usepackage{amsmath}
\usepackage{bbm}
\usepackage{float}
\usepackage{url}

\newcommand{\real}{{\mathbb R}}

\newcommand{\captionfonts}{\footnotesize}
\makeatletter
\long\def\@makecaption#1#2{%
  \vskip\abovecaptionskip
  \sbox\@tempboxa{{\captionfonts #1: #2}}%
  \ifdim \wd\@tempboxa >\hsize
    {\captionfonts #1: #2\par}
  \else
    \hbox to\hsize{\hfil\box\@tempboxa\hfil}%
  \fi
  \vskip\belowcaptionskip}
\makeatother 
\begin{document}
\title{Relativity Theory Refounded}

\author{Diederik Aerts \vspace{0.5 cm} \\ 
        \normalsize\itshape
        Center Leo Apostel for Interdisciplinary Studies, 
         Brussels Free University \\ 
        \normalsize\itshape
         Krijgskundestraat 33, 1160 Brussels, Belgium \\
        \normalsize
        E-Mail: \url{diraerts@vub.ac.be}
           \\
              }

\date{}
\maketitle
\begin{abstract}
\noindent
We put forward a new view of relativity theory that makes the existence of a flow of time compatible with the four-dimensional block universe. To this end, we apply the creation-discovery view elaborated for quantum mechanics to relativity theory and in such a way that time and space become creations instead of discoveries and an underlying non-temporal and non-spatial reality comes into existence.
We study the nature of this underlying non-temporal and non-spatial reality and reinterpret many aspects of the theory within this new view. We show that data of relativistic measurements are sufficient to derive the three-dimensionality of physical space. The nature of light and massive entities is reconsidered, and an analogy with human cognition is worked out.
\end{abstract}
\begin{quotation}
\begin{center}
{\bf Preface} 
\end{center} 
\bigskip
\noindent
She received me with warmth and led me to the dance floor in the center of the ballroom. I told her how happy I was to see her. She smiled and showed me a place at one of the tables. She asked me what I would have, adding that there were many tasty dishes. And then, in a more serious voice, she told me there was a man waiting to see me.  ``The gentleman seems intent on meeting you,'' she said, with a frown on her forehead, ``he asked me several times whether you had arrived yet". Then she smiled again, ``I will go and tell him that you are here, and also fetch you something to eat", and she hurried away. I saw her disappear into the crowd on the other side of the ballroom. I was sitting at the table, watching the dancing couples. Not much later she came back with all sorts of goodies for me. By her side I saw Albert Einstein, clearly recognizable by his tousled grey hair and wrinkled face. ``This is the gentleman who wants to meet you,'' she said as she led him to the place at the table right in front of me. Einstein greeted me kindly and while I was still in great amazement, we began a conversation that was to last until late in the night.

\bigskip
\noindent {\it Dedicated to the cherished and sweet memory of Natalie}
\end{quotation}
\bigskip
Unlike most approaches, we will not attempt to build relativity theory \cite{einstein1905,einstein1905b,minkowski1915,einstein1916,einstein1920,einstein1952,misnerthornewheeler1973} from as small a set of axioms as possible.
Although such an axiomatic construction is very valuable, we believe that it keeps at least some of the essential aspects of `understanding' hidden, because of the excessive focus on the physical content of the specific axioms that constitute a minimal set. By contrast, our presentation and analysis of relativity theory will focus from the start on those properties of the physical entities under consideration that are `intrinsically real' or, using the standard terminology of relativity theory, that are `proper'. Further, we will make use of the mathematical structure of the theory to model specific situations and rely on the fact that this structure has been tested experimentally in innumerable ways that have proved to provide faithful models of such situations. We will also be quite frank in some of the aspects of our view and not let ourselves be misguided by new taboos arising mainly due to the intellectual struggle associated with the historical development of relativity theory -- we allow ourselves to speak about `the flow of time', for example. Of course, we will have to show  that such a notion as `the flow of time' makes sense, i.e. that we can understand its meaning within our view on relativity theory. We will also make explicit use of knowledge about the nature of reality derived from quantum theory in general and also from our conceptual interpretation of it \cite{aerts2009,aerts2010a,aerts2010b,aerts2013,aerts2014}. As we shall see, it is in a conversation with relativity theory as it is usually perceived, taking into account insights from quantum theory and also being open to intuitions beyond, which allows us to put forward a new view of the physical reality described by relativity theory, giving rise to a possible understanding of the deep problematic conceptual problems of interpreting the nature of this physical reality.

\section{Intrinsic aspect of physical entities}

An important difference with the traditional view on relativity is that we will consider physical entities not a priori as `material object occupying a specific time-space region inside a universe that is also inside time-space'. We rather want to consider time-space as a `theatre of encounter for physical entities'. This is the reason why we make an attempt to build the theory starting as much as possible from what we have learned from traditional relativity theory -- its classical framework as well as the known quantum versions -- are the intrinsic aspects of such a physical entity.

As we will see, in our view, for a specific physical entity the flow of time exists, and contrary to what is often believed, this flow of time is intrinsic. It is what is called in relativity theory `the proper time of a physical entity'. Using the notation for proper time in relativity theory, we denote this time by the Greek letter $\tau$. This time can be measured by any sufficient regular repetitive part of the physical entity we consider, let's call such a repetitive part a clock. A physical entity, if not a point particle, has an extension, i.e. takes a place while it also takes a time. Also this extension can be given an intrinsic measure,  depending on the specific geometric structure of the physical entity, it can be specified by giving the sizes of the lengths separating different points needed to characterize this geometric structure. We measure these lengths by what in relativity theory is called the `proper length'. We will denote this length, taking over the common notation for relativity theory, by $s$. In fact, as long as we consider one unique specific physical entity, not much more can be said, it has an intrinsic extension -- including a specific geometric form, and an intrinsic elapsing of its time.

Of course, reality consists of more than one physical entity, hence let us consider two such entities $A$ and $B$, both equipped with clocks that measure their respective times $\tau^A$ and $\tau^B$, and also equipped with two extensions measured by lengths $s^A$ and $s^B$. We want to start considering the simplest of all imaginable situations, but one for which we know that it exists in our reality, which is the following. Both physical entities $A$ and $B$ are at rest with respect to each other. We have not yet specified anything about the nature of the physical entities, and will do so now, just to make the situation easier to imagine. So let us suppose that both entities are pieces of matter with masses $m^A$ and $m^B$ and we indicate by $(x_1^A, x_2^A, x_3^A)$ and $(x_1^B, x_2^B, x_3^B)$ the center of mass for each physical entity as the point to indicate its place -- by introducing three numbers to indicate these centers of masses, we already prelude the fact that we will start considering more complex situations where at least one of the physical entities moves with respect to the other, and kind of also prelude that there are three orthogonal dimensions describing how such a movement can take place, but these are not essential elements, and rather introduced here to keep our analysis simple enough to be able to focus on the core of the matter. 

Hence, as we said, $A$ and $B$ have places indicated by points $(x_1^A, x_2^A, x_3^A)$ and $(x_1^B, x_2^B, x_3^B)$ and these points do not move with respect to each other -- and this is also the case for any other two points we would have chosen to identify the places of $A$ and $B$. Another way of stating the same would be to say that $A$ is in rest with respect to $B$ and $B$ is in rest with respect to $A$. We know from experience that this is possible -- we will later see that when the effects described by general relativity become important, we have to be more careful in defining well this type of situation, but for the time being, this is how we start. In such a situation, $A$ and $B$ can easily synchronize their clocks, and from relativity theory follows that if they do so, both clocks measure the same intrinsic time. This means that we can put in this situation $\tau^A=\tau^B$. Also for the measuring of length, $A$ and $B$ can use the same measuring rod, and agree about all necessary aspects of the procedure of length measuring, and again, relativity theory teaches us that $s^A=s^B$ in this case of $A$ and $B$ being at rest with respect to each other. This means that $A$ and $B$ can measure an absolute distance that separates them. Let us make very clear, however, at this stage, because here the customary confusion in most interpretations of relativity theory already starts, `what we have identified as being intrinsic' and also `what we have not identified at all'. We have identified that both $A$ and $B$ have the same intrinsic flow of time, hence $\tau^A=\tau^B$, and they can synchronize their clocks, which will remain synchronized as long as $A$ and $B$ remain in this way at rest with respect to each other. We have also identified that $A$ and $B$ can measure with the same intrinsic length, hence $s^A=s^B$, and in this way can determine an intrinsic length that separates them. What we have not done is identify a time-space where both are present, such that their time would be the time elapsing in this time-space and their lengths would be measured by the distances of this time-space. This, i.e. the identifying of such a time-space, we have `not' done. And, as we will see in the analysis we will make now, it is something we should only do with extreme care.

The next situation we want to analyze, taking into account our knowledge of relativity theory, theoretical as well as experimental, is where one of the two physical entities starts to move with respect to the other. Let us say that $B$ starts to move with respect to $A$, and $A$ remains without moving -- we will come later to specifying more carefully still what we mean by this. To make the calculation simpler, we suppose that $B$ moves with constant velocity equal to $0.9c$, where $c$ is the velocity of light, in the direction $x$.
Let us make still more explicit what we mean exactly by this. We mean that the intrinsic distance measured by $A$ -- hence the intrinsic distance for $A$, and not for $B$ -- towards $B$ increases in size with the time flow of $A$ -- hence the flow of the intrinsic time of $A$ and not of $B$ -- in a constant way, for example in one second the distance has increased 0.9 light second, a light second being the distance traveled by light in 1 second. Relativity teaches us that the `time flow' of the moving $B$ -- hence the intrinsic time of $B$ and not of $A$ -- will slow down substantially, and we can exactly calculate how it slows down. Let us become concrete, and suppose that $A$ remains without moving from a time $\tau^A=0$, to a time $\tau^A=10$ years. In these 10 years, moving with a velocity equal to $0.9c$, $B$ has moved away from $A$ over a distance $\Delta x=9$ light years. 
Then the time $\tau^B$ will have elapsed the amount equal to 
\begin{eqnarray}
\Delta \tau^B=\sqrt{{\Delta \tau^A}^2-({v\Delta \tau^A \over c})^2}=4.3589\ {\rm years}
\end{eqnarray}
So, less than half of the time has elapsed for $B$ as compared to the time that has elapsed for $A$.

Immediately two questions arise. The first question, the one of the famous twin-paradox, is the following. Why would we not look at the situation the other way around, and consider $B$ to be not moving while $A$ is moving in the opposite direction? If we were allowed to make this `relativistic' symmetry reasoning, we could as well come to the conclusion that time has elapsed more slowly for $A$ instead of for $B$. A first -- but, as we will see, incorrect -- answer would be the one that comes to mind right away, namely the following. Only one of the two `starts' to move, and hence undergoes during a certain time an acceleration for speeding up to the constant velocity, and the other, since remaining at rest, does not experience this acceleration. The presence of this acceleration breaks the symmetry of the situation. We can, however, making the situation slightly more complex, overcome the necessity of having an acceleration involved. Indeed, suppose that $B$ was moving all the time with constant velocity, and both $A$ and $B$ just take the encounter as an opportunity to synchronize their clocks, in principle this is possible, and then the situation is indeed completely symmetric. For such a situation, where $B$ was already moving, we indeed can equally well decide that $A$'s clock is slowing down, or even that both clocks are behaving still in a more complicated way with respect to each other. We will come back to this in detail after a further analysis, which we will make right away.

Indeed, we can only decide about the problem just mentioned if we configure further the situation such that $A$ and $B$ meet again in their common future. Suppose this happens by `only' $B$ having to stop moving away from $A$, turning around, and to start moving towards $A$ with the same velocity. Then, after 20 years have passed for $A$, they will meet again, and only 8.7178\ {\rm years will have passed for $B$. Does the asymmetry comes now from the fact that $B$ needs to experience an acceleration during the turning around? Again, this is not the case. We could involve a third physical entity $C$, which encounters $B$ exactly after 4.3589 years have passed for $B$, synchronizes clocks, and then moves towards $A$ with the inverse velocity of $B$. Then $C$ will encounter $A$ after 4.3589 years exactly, carrying a message from $B$ who left $A$ 8.7178 years ago on $C$'s clock. And on $A$'s clock 20 years will have passed. So, it is not the effect of the presence of acceleration which causes the `slowing down of time flow'. It is the geometric structure of the paths crossed by $A$, $B$ and $C$ which contains the fundamental asymmetry which is predicted by relativity theory and also has been observed on numerous occasions experimentally.

By the way, the above analysis also makes clear that it is not correct to interpret the difference between the passage of time for $A$ as compared to $B$ -- suppose for a moment we forget again about $C$, and hence are considering the situation where $B$ turns around and after the turn heads back towards $A$ -- as due to a physical effect that, for example, would be forced upon the mechanics of the clock of $B$ as a consequence of the acceleration during the turning around. There is no `physical mechanical effect on clocks' involved in the time dilatation effect of relativity theory. The difference between $\tau^A$ and $\tau^B$, and let us repeat that `both' are intrinsic times for the physical entity $A$ and $B$, respectively, is due to $A$ and $B$ having travelled a different path in the time-space structure of relativity theory. Moving, what $B$ does with respect to $A$, does not only give rise to `moving in space', but also simultaneously to `moving in time'. Well, if only for a moment, in what we just wrote, we have given in to the desire to consider a global time-space structure in which $A$ and $B$ would be present and could move around. Let us give in to this desire deliberately now, such that we can see which are the deep paradoxes that arise from it.

If there is a time-space continuum, for special relativity -- which contains the core of the mystery -- this would be the Minkowski time-space, in which four vectors are intrinsic entities -- then, as we concluded already above, the difference in elapsed time between a moving physical entity such as $B$ and an entity at rest such as $A$, is due to $B$ and $A$ taking another path through time-space to travel from the event where they first encounter to the event where they encounter for the second time. The path of $A$ is a path that connects the two time-space events $(0, x_1^A, x_2^A, x_3^A)$ and $(20\ {\rm years}, x_1^A, x_2^A, x_3^A)$ by means of a straight line parametrized as $(x_0^A, x_1^A, x_2^A, x_3^A)$, where $x_0^A$ runs from $0$ to 20 years. The intrinsic time $\tau^A$ which passed for $A$ is calculated from the general formula of Minkowski time-space given by
\begin{eqnarray}
\Delta \tau=\sqrt{(\Delta t)^2 - ({\Delta x_1 \over c})^2 - ({\Delta x_2 \over c})^2 - ({\Delta x_3 \over c})^2}
\end{eqnarray}
Given that for $A$ we have $\Delta x_1$=$\Delta x_2$=$\Delta x_3$=0 and $\Delta t$= 20 years, we get
\begin{eqnarray}
\Delta \tau^A=\sqrt{({\rm 20\ years})^2}={\rm 20\ years}
\end{eqnarray}
The path of $B$ is a very different one. It first connects, also in a straight line, the points $(0, x_1^B, x_2^B, x_3^B)$ and $(10\ {\rm years}, x_1^B+9\ {\rm light\ years}, x_2^B, x_3^B)$, and then, after $B$ has turned around to head again towards $A$, it connects, again in a straight line, the points $(10\ {\rm years}, x_1^B+9\ {\rm light\ years}, x_2^B, x_3^B)$ and $(20\ {\rm years}, x_1^B, x_2^B, x_3^B)$. If we calculate by means of the same intrinsic definition for time in Minkowski space the time elapsed for $B$ on these two paths, we find, for both paths $\Delta x_1$=9 light years, $\Delta x_2$=$\Delta x_3$=0, and $\Delta t$= 10 years. Hence the time elapsed on each path is given by
\begin{eqnarray}
\Delta \tau^B\ {\rm (each\ path\ for}B{\rm )}=\sqrt{({\rm 10\ years})^2 - ({{\rm 9\ light\ years} \over c})^2}=4.3589\ {\rm years}
\end{eqnarray}
hence a total time elapsed of the double, which gives
\begin{eqnarray}
\Delta \tau^B\ {\rm (total\ path\ for}B{\rm )}= 8.7178\ {\rm  years}
\end{eqnarray}
So, indeed, we see that the difference is due to the difference in path taken by $B$ as compared to $A$ to encounter each other in two events $(0, x_1^A, x_2^A, x_3^A)$ and $(20\ {\rm years}, x_1^A, x_2^A, x_3^A)$ for the case of $A$, and $(0, x_1^B, x_2^B, x_3^B)$ and $(20\ {\rm years}, x_1^B, x_2^B, x_3^B)$ for the case of $B$. And, this difference is due to the structure of Minkowski time-space as a four-dimensional manifold.

Hence, we have to conclude, at this point of our analysis, that the geometric interpretation of relativity holds, namely that `time-space' is what intrinsically exists, while `time' and `space' as separated entities cease to exist. Customarily this is called the block universe interpretation of relativity. It comes along, however, with severe trouble. In case reality `is' the four-dimensional time-space of Minkowski, what is then the meaning of `change in time'? Does this mean that physical entities exist within the four-dimensional time-space manifold, and are their world-lines? And when a human being experiences a physical entity, does he or she experience a `slice' in the time-space continuum of this physical entity's world line? But, if this was the case, why are we as individuals not four-dimensional? We definitely are not our past and future all at once? Does this then mean that it is our consciousness that in some way `travels' on the world-line which our body is, our body being four-dimensional? Would this also mean that for what concerns physical reality change is not possible, and is only an illusion provoked by our consciousness, while the future is as fixed as the past and the present?

Or, do we have to go to the other extreme, and is existence not four-dimensional at all? And is Minkowski time-space only a mathematical construction, and travel for all physical entities the entities on their world-lines? The problem with this view is that, and this follows from the reasoning we described above, the relativistic effects really come about as a consequence of the different paths connecting time-space points in the four-dimensional Minkowski manifold. They cannot be retrieved as due to effects on individual physical entities that travel on paths, because they are due to the way such a path is part of the global four-dimensional manifold. Hence, we can say that the matter is truly characterized by something related to `four dimensions', and that any attempts to escape this are bound to fail.

Opinions are divided about these two options, and nobody, as far as we know, seems to really understand the situation. We believe that the reason is that both views are wrong, and it is a third view, quite different from the two ones that are encountered in the literature which is the correct one. It is this third view that we want to start to elaborate here.

\section{The reality beneath space-time}
The view that we propose needs insight from quantum mechanics, and hence relies on the fact that relativity theory without quantum mechanics is not complete. More specifically, it relies on the creation-discovery view that we developed for quantum mechanics and the hypothesis to interpret quantum non-locality as quantum non-spatiality \cite{aerts1998,aerts1999}. We go a step further, however, and put forward the hypothesis that physical entities are intrinsically non-spatial and also non-temporal. And that `time' and `space' are consequences of a creation aspect within our creation discovery view.

Of course, like the saying goes, `the proof of the pudding is in the eating', and hence we have to try to show that the above-mentioned hypothesis makes sense. How better to provide such proof than by giving a simple example that enables us to can see and understand what could be the matter for physical reality? Our example is inspired by our `conceptual interpretation of quantum mechanics' \cite{aerts2009,aerts2010a,aerts2010b,aerts2013,aerts2014}, although this does not mean a priori that it is also an argument in favor of this interpretation. Here, we only want to show how the view on relativity theory -- and hence also quantum mechanics with non-spatiality and non-temporality -- that we put forward here can be true, and what are some of the immediate consequences in case it is true.

Our example consists in considering a definitely non-spatial and also non-temporal collection of entities, namely the conceptual meaning structure of humanity. In a first stage, to be as concrete as possible, we consider it in the form of the World-Wide Web. How do we find a time-space structure connected to the World-Wide Web? Well, each time that we log in to the World-Wide Web using our computer and our browser, and we see in front of us a specific webpage, we consider this as `a place', where the meaning content of that webpage `takes time and place', hence `becomes localized in time and space'. Indeed, also a specific instant of time, for example, connected to the click of our mouse on the link given to us by Google after a search, and the opening of the webpage, is connected to it. If we push on a specific link that we see on the webpage, we move to another webpage, and also to another meaning content which at that time gets localized on the newly appeared webpage. We could have pushed another link, and this would have brought us to usually another webpage, possibly to the same, however. Anyhow, we all know very well the dynamical process I put forward here, it is called `surfing the World-Wide Web'. Suppose we consider two persons $A$ and $B$ surfing and starting from the same webpage with a mutual experience that they `meet before starting both their surfing path', and such that, after having taken a different path through the World-Wide Web, also ending up at the same webpage in a mutual experience that `they meet again'. So they started together and meet again together. Quite obviously, if we measure intrinsic time for both with the repetitive actions of their clicking of new links, this time in general will be quite different for $A$ and $B$, and exactly depending on the paths that both have followed to get from the first starting webpage to a second webpage were they meet again. The difference in time is intrinsically due to the structure of the World-Wide Web, and the different paths that can be taken to go from one webpage, were $A$ and $B$ meet in the beginning, to another webpage, where $A$ and $B$ meet the second time.

That the foregoing scenario is realistic, i.e. it can be realized, and counting the number of clicks that both paths need to meet again at the same webpage, we can easily calculate the difference of the flow of time for $A$ and $B$. The reason that it can happen is because the World-Wide Web represents a coherent collection of meaning. We can add some additional aspects to give our example more explicative power. For example, every webpage could have some links that direct to the same webpage. If both $A$ and $B$ proceed with these links, they will follow two paths with times that are equal, at least if they synchronize their clicking speed -- but that is also necessary in physical reality. Pushing another link, in the analogy, would mean already `no longer being at rest with respect to each other'. It will be completely determined by the structure of the underlying non spatial and non temporal World-Wide Web, how the different paths of $A$ and $B$ will fare in between them reaching the same webpage again at the same time, and hence meeting again. But, the time that elapses for both will be intrinsic, because totally determined by the structure of the underlying non spatial and non temporal World-Wide Web. Namely the time will be completely determined by the number of webpages lying in between , the number of places that are encountered in between to move over this specific path from the first meeting spot to the second meeting spot.

Can we understand by means of this example that we do not get into the trouble of the block universe interpretation? Indeed, it follows from the example that there is a real past, which has passed, and there is a present -- although only locally for one entity -- and there is a future which is not fixed. What is however given and existing outside of the time measured by the clicking of surfers is the meaning structure of the World-Wide Web. This is not a four dimensional time-space structure, because it is non spatial and non temporal, but its realization as a consequence of surfing it, does get related in a specific way to such a four dimensional manifold structure. That also the `lapsing of time' gives rise to a dimension which is structured as if it were a space dimension is also the case for our example, but does not mean that this time-dimension `is' a real existing piece of reality. The structure of the time-dimension follows from the structure of the non temporal and non spatial World-Wide Web, and indeed `how the space dimensions get structured' is equal to `how the time dimension gets structured', while what is underlying is always the structure of `how meaning is coherently connected' - we will analyze later how `space appears for the example of the World-Wide Web'.

If we apply this explanation to physical reality, does this still lead to physical entities being four dimensional world-lines? No, but it does mean that physical entities stick out their coherent nature into their future and their past too, like they do in the space directions  - and a space direction means for a physical entity `those places where another physical entity is'. But `it is not an already determined future' while `it is a past that has passed'. Hence, it is not in the way that `they exist in the future', but only due to the fact that observers use links that were already inside the meaning structure of physical reality when making their path to the future. But the parts of the already existing reality, which is non temporal and non spatial, like the meaning content of the World-Wide Web is, has no `time dimension' connected to it, `before the surfing action has chosen which path to take. If we define `happenings' as `meaning parts of the World-Wide Web -- hence no surfing yet involved -- then the `taking time and place' is part of the `creation' part of the creation-discovery dynamics. Time-space is not discovered but created, but a vast underlying non temporal and non spatial structure `is' present, in our example it is the meaning structure of the World-Wide Web. Let us also mention that the World-Wide Web in itself also changes constantly, but this change is not related to the change taking place while surfing. Hence, this could be the same for physical reality, which changes constantly, but this change is not essentially related to the change we see if we surf it wearing space-time clothes.

It is, by the way, here that we would like to add an aspect that is not present in the World-Wide Web, but that would have been present if we had considered a deeper version of the meaning structure of the human realm. In the paths followed by physical entities also direct creation of meaning is possible. In the example of the World-Wide Web only already existing meaning and existing links can be followed. Hence, in this respect the example should be enriched to capture the more quantum nature of physical reality, and then the two types of change mentioned above would be linked. But that was not the aim of it here, the aim rather being to show that with this example we can understand relativity theory.

Let us analyze what the equivalent of `space' is within this view. The `space' surrounding a physical entity is the collection of `places' where other physical entities can be together with the considered physical entity, and we use here `being` as non temporal and non spatial being, where `temporal' relates to the `time elapsing for each individual physical entity' while surfing. 

Without being very explicit about it, we already used aspects of the notion of space when we reasoned about $A$ and $B$ being `at rest' with respect to each other. We have to try to see clear in these notions now for our example of the World-Wide Web. Suppose we consider a webpage $A$ that we open on our computer, then obviously there is an enormous amount of other webpages that `we could have opened' but we did not. They all are webpages that we could have opened if we had made another decision in the past of the procedure of opening actually webpage $A$ \cite{aerts1996a,aerts1996b}. Also, someone else could actually be looking at the same webpage simultaneously and then surf in another direction. How to go about to introduce more specific aspects of these `space-like' situations and do this in such a way that the comparison with the situation of physical entities in Minkowski space can be meaningfully made. We proceed as follows. 

We suppose that `surfing' is continuously happening. So, when we consider a snapshot of it, for example, webpage $A$ that is looked at, we suppose that the webpage was realized by clicking a link of another website, that was open `before' $A$ was open. And then it continues, after $A$ being open, another webpage will be open by clicking a link on $A$. Let us also mention here that in our example, surfing happens actively due to a human being, however it is not Einstein's `observer' that the role of the human being presents in case we go from our example to the situation of the physical world. It are physical entities themselves that play the equivalent of the role of surfing, and that hence `create' parts of time-space when being as they are. That is why in the opening text of this investigation we mentioned that we will put forward time-space as a theater of encounter of physical entities.

Next to single webpages, we will introduce world lines which are `series' of webpages $(A_n)_n$, where $n$ runs from $0$ to the natural number $m$. So, more concretely, the  world line $(A_n)_n$ consists of the set $\{A_0, \ldots, A_m\}$ of webpages surfed through one after the other, and the world line $(B_k)_k$ consists of the set  $\{B_0, \ldots, B_l\}$ of webpages, surfed through one after the other.

Let us consider $A_0$. For the concrete world line $\{A_0, \ldots, A_m\}$, the webpage $A_0$ contains numerous other links that could have been pushed instead of the link that leads to the realization of the world line $\{A_0, \ldots, A_m\}$. Let us suppose now that $(A_n)_n$ and $(B_k)_k$ start of at the same webpage $A_0=B_0$ but with other links being pushed which means that they go off in different directions after starting from the same webpage. A possible way is that a second person $B$ surfs through the world line $\{B_0, \ldots, B_l\}$, and starts his or her surfing together with person $A$ at the same webpage. Additionaly we suppose that both also end at the same webpage, hence $A_0=B_0$ and $A_m=B_l$. In general $m$ will be different from $l$, and if we consider $m$ and $l$ respectively as measures of the intrinsic times that passed by when $A$ and $B$ ran through their respective world lines, this would mean that different intrinsic times passed by for $A$ and $B$. We can consider the set of all world lines that start at the same webpage $A_0$ and end at the same webpage $A_m$, and let us call this set ${\cal A}_{0,m}$. Without loss of generality we suppose that $(A_n)_n$ is the longest of all the world lines of ${\cal A}_{0,m}$, or with other words that all $l$'s for all other elements of ${\cal A}_{0,m}$ are smaller or equal to $m$.

\section{Minkowski coordination}

We are now ready to introduce a Minkowski type coordination for ${\cal A}_{0,m}$. We proceed as follows. We choose a time-axis that coordinates the elements $A_n$ of the world line $(A_n)_n$ at equal spaced points on the axis, and without loss of generality, since it consist only of fixing a unit time interval, we choose the coordinate numbers $0, \ldots, m$ for the elements of $(A_n)_n$. Let us consider now the world line $\{B_0, \ldots, B_l\}$ element of ${\cal A}_{0,m}$. We know that $B_0=A_0$ and $B_l=A_m$, which means that we will coordinate $B_0$ and $B_l$ by the same points $0$ and $m$ of the introduced time axis. What about the other webpages, starting with $B_1$ and continuing to $B_{l-1}$? It is here that `space' comes into being as a way to give a place to webpages that exists together with the ones that we have coordinated already on the time axis. Of course, at once this means that we also give a place to the elements of $(A_n)_n$, namely, `they are at rest in the origin of the coordinate system that we put up'. This means that the elements of $(B_k)_k$ will not be at rest, but moving. But, all this needs to follow from our careful operational construction, so let us proceed step by step.

An arbitrary element of $(B_k)_k$, for example the element $B_k$ will be given a time-space coordinate indicating its place as an element of the set ${\cal A}_{0,m}$ and of the time-space coordinate system that we introduce, in the following way. Consider the interval $[t_0(A), t_m(A)]$ where, $t_0(A)$ and $t_m(A)$ are the time coordinates of $A_0$ and $A_m$. Then all other time coordinates of elements of $(A_n)_n$ are points inside $[t_0(A), t_m(A)]$ of equal distance from each other. We choose $t_0(B)=t_0(A)$ and $t_m(A)=t_l(B)$, and divide the interval $[t_0(A), t_m(A)]$ in $l$ equal parts to give time coordinates to all other points of $(B_k)_k$. However, contrary to all elements of $(A_n)_n$ having space coordinate equal to zero, this is no longer the case for coordinates of the elements of $(B_k)_k$ different from $B_0$ and $B_l$. Let us explain how we coordinate all points of $(B_k)_k$. To $B_0$ we give coordinate $(t_0(B), 0)=(t_0(A),0)$, expressing that both $(A_n)_n$ and $(B_k)_k$ take off as world lines at the websites $A_0=B_0$ located both at coordinate $(t_0(A), 0)=(t_0(B), 0)$. Without loss of generality, we can choose $t_0(A)=t_0(B)=0$, which consists of letting the origin of the time-space coordinate system be at the beginning webpage for both world lines.

We introduce a positive number $c$, which will play the role of the velocity of light in our example, and analyze later its meaning. For webpage $B_1$ we choose the coordinate 
\begin{eqnarray}
(t_1(A), c\sqrt{m^2-l^2 \over m^2})
\end{eqnarray}
Remembering the Minkowski metric and it being invariant, we can interpret this choice of time-space coordinate for $B_1$ as follows. In the reference frame $(t, x)$ that we introduced, and where $A$ is at rest, $B$ has moved from $(0,0)$ to $(t_1(A), c\sqrt{m^2-l^2 \over m^2})$ in a time interval $[0, t_1(A)]$.This means that a velocity 
\begin{eqnarray}
v={c\sqrt{m^2-l^2 \over m^2} \over t_1(A)}
\end{eqnarray}
is involved. Again without any loss of generality we can choose $t_1(A)=1$, which comes to taking $t_1(A)$ as our time unit in the considered reference frame. This means that $B$ is moving away from $A$ in the positive direction of the space axis $x$ with velocity $v=c\sqrt{m^2-l^2 \over m^2}$.

Let us analyze this. The fraction of the velocity of light $c$ that determines the velocity $v$ is given by ${\sqrt{m^2-l^2} \over m}$. This means the following. When a link is chosen, then ${\sqrt{m^2-l^2} \over m}$ stands for the velocity through space, i.e. expressed as fraction of the velocity of light, which is the maximum velocity through space which is possible, that this links carries with itself related to the purpose of `reaching the end webpage', where meeting is taking place.

Let us look at two extremes. When $l=m$, then this velocity is zero. Indeed, it means that the chosen road is equally slow as the slowest one, which is $(A_n)_n$. Since we directed our time axis along this one, space is only involved in a passive way, i.e. no movement is space takes place. Consider the other extreme, namely $l=0$. This means that `no link is needed to go from the begin webpage to the end webpage'. In fact, this is a limit situation not possible to reach by surfing, where the fastest way would be `one click', hence `one link'. But anyhow, in this limit situation of zero links, we get that $v=c$, namely that we need to move through space with a velocity equal to that of light.

However, we wonder whether we really want to represent $B_1$ by the time-space coordinate $(t_1(A), c\sqrt{m^2-l^2 \over m^2})$. We know that $t_1(A)$ is the moment in time when the click is made to surf from $A_0$ to $A_1$, but this is not the moment in time when the click is made to surf from $B_0$ to $B_1$. Hence, we should represent $B_1$ by a time-space coordinate where the time coordinate is $t_1(B)$ and not $t_1(A)$. And indeed, that is what we will do. Let us calculate this time-space coordinate. We have $t_1(B)={m \over l}t_1(A)$ and hence the coordinate that we look for is given by 
\begin{eqnarray}
(t_1(B), {m \over l}c\sqrt{m^2-l^2 \over m^2})=({m \over l}, c{\sqrt{m^2-l^2} \over l})
\end{eqnarray}
 when we have put $t_1(A)=1$.

Let us construct now the coordinates of all the elements of $(B_k)_k$. We make the additional hypothesis that both $m$ and $l$ are even numbers -- the uneven case needs a slightly different treatment, but is essentially completely analogous, hence we leave its details to work out for the interested reader. The following are then the time-space coordinates of all elements of $(B_k)_k$.
\begin{eqnarray}
&&B_0 \leftrightarrow (0, 0) \\
&&B_1 \leftrightarrow ({m \over l}, c{\sqrt{m^2-l^2} \over l}) \\
&&B_2 \leftrightarrow 2({m \over l}, c{\sqrt{m^2-l^2} \over l}) \\
&&\ldots \nonumber \\
&&B_{{l \over 2}-1} \leftrightarrow ({l \over 2}-1)({m \over l}, c{\sqrt{m^2-l^2} \over l}) \\
&&B_{l \over 2} \leftrightarrow ({m \over 2}, c{\sqrt{m^2-l^2} \over 2}) \\
&&B_{{l \over 2}+1} \leftrightarrow  ({m \over 2}, c{\sqrt{m^2-l^2} \over 2})+({m \over l}, -c{\sqrt{m^2-l^2} \over l}) \\
&& \dots \nonumber \\
&&B_{l-1} \leftrightarrow (m-{m \over l}, c{\sqrt{m^2-l^2} \over l}) \\
&&B_l \leftrightarrow (m, 0)
\end{eqnarray}
In Figure 1 we have graphically illustrated the situation for $m=10$ and $l=8$.
\begin{figure}[htbp]
\begin{center}
\includegraphics[scale =1]{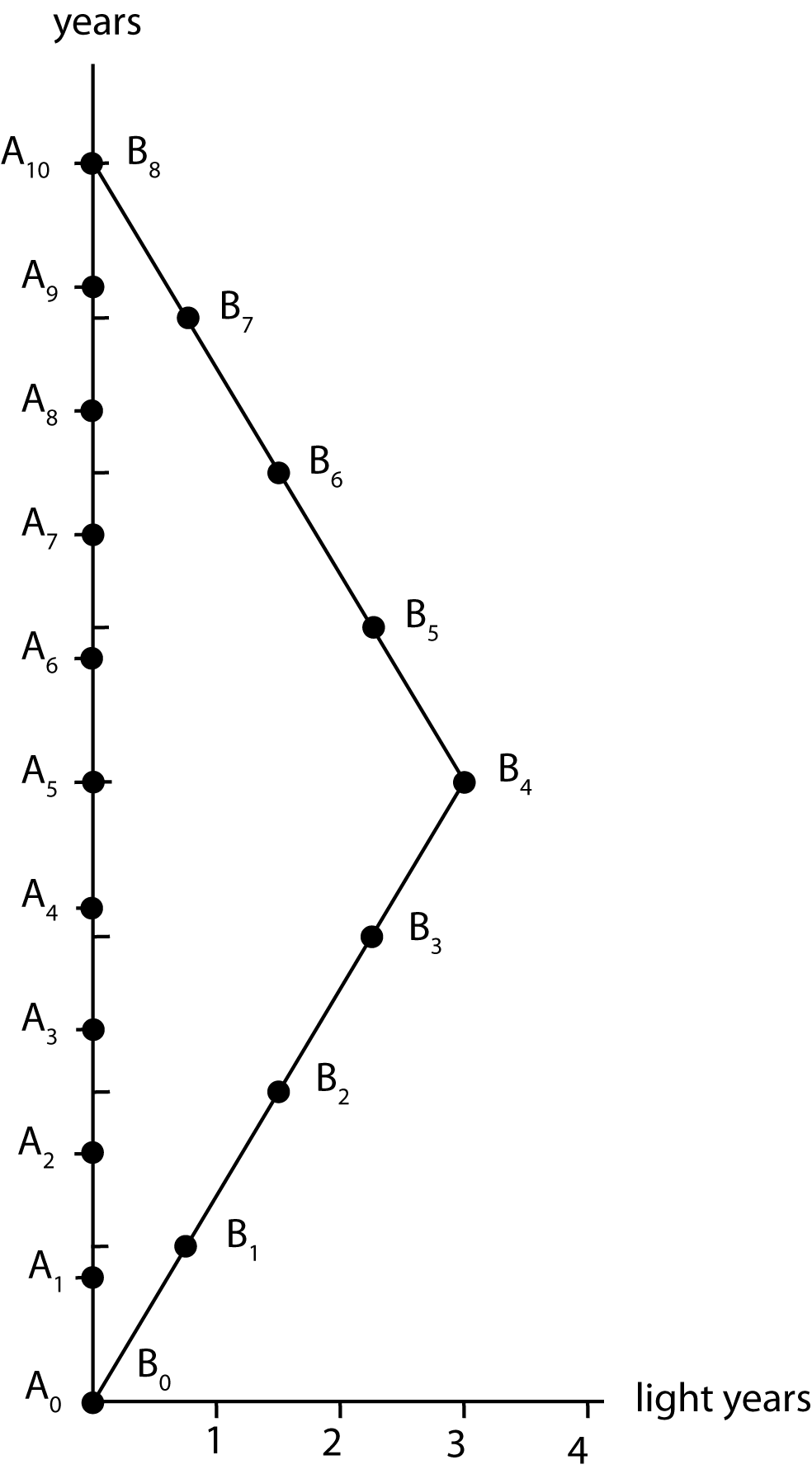}
\caption{A graphical representation of $A$ and $B$ for $m=10$ and $l=8$}
\end{center}
\end{figure}
and with units of time `years' and unit of length `light years'. Hence, in the situation represented in Figure 1, we have
\begin{eqnarray}
&&B_0 \leftrightarrow (0, 0) \quad B_1 \leftrightarrow ({5 \over 4}, {3 \over 4}) \quad B_2 \leftrightarrow ({5 \over 2}, {3 \over 2}) \quad B_3 \leftrightarrow ({15 \over 4}, {9 \over 4}) \quad B_4 \leftrightarrow (5, 3) \\
&&B_5 \leftrightarrow ({25 \over 4}, {9 \over 4}) \quad B_6 \leftrightarrow ({15 \over 2}, {3 \over 2}) \quad B_7 \leftrightarrow ({35 \over 4}, {3 \over 4}) \quad B_8 \leftrightarrow (10, 0)
\end{eqnarray}
Like can be seen on Figure 1, we have constructed the world line for $B$ in such a way that $B$ first moves with velocity $v={\sqrt{m^2-l^2} \over m}$ in the positive direction of the $x$-axis, hence moves `away' from $A$. Then halfway -- which is the reason that we introduced the hypothesis for $m$ and $l$ to be even numbers -- $B$ turns around, and starts to move with the same magnitude of velocity $v={\sqrt{m^2-l^2} \over m}$, but in the opposite direction, hence approaching $A$ again. This is the reason that both can meet again at $A_m=B_l$. Meanwhile however $m$ years have passed by for $A$ -- or $A$ has clicked $m$ webpages while surfing -- and only $l \le m$ years have passed by for $B$ -- or $B$ has clicked $l$ webpages while surfing. Everybody can easily recognise the typical situation considered in the twin paradox of special relativity theory.

There are different reflections to be made. First of all, the representation that we gave for $(B_k)_k$ is not general, there are many different configurations possible that realise within a Minkowski metric the situation $l \le m$. What is however interesting to remark is the following. In case we limit ourselves to `constant velocities and movements in straight lines' -- possibly with making a turn around brusquely like in the case of our example -- the different possibilities all come to the same in principle. It is always also necessary to make a turn around, if we start by representing the slowest path by a straight line without turning around. What is the meaning of this? We have to invoke general relativity perhaps to be able to understand better. Indeed, instead of straight line and constant velocity and a brusque turn around, it would be possible to realise the path of $B$ by means of a curved line, were accelerations play a role. And, one could switch completely to general relativity. The slowest path -- hence $A$ -- would then be the path following a geodesic of the gravitational situation. It is more easy to understand that `this geodesic' path is the one we should give preference in coordinating, of course, in the case of general relativity the coordination would only locally be Minkowski. What need to be mentioned is that even then the counter intuitive aspect -- which we will try to formulate better and analyse deeper right away -- remains present, because locally Minkovski metric governs, which means that time-space is curved hyperbolically.

This hyperbolic curvature of Minkowski space means that `velocity, even if we identify it at first hand as a movement through space' is mainly a movement through time. However, like our example of the World-Wide Web and surfing shows, this does not mean that we need to go towards a block interpretation of reality. What does it mean then? Well, it means that we can influence the way we reach out in the future by getting ourselves moving, and if we accelerate the influence becomes even bigger -- we literally move faster and a time coordination will reveal this. If we let ourselves float on a geodesic we move the most slow way possible. Movement is however `over the different parts of the underlying non temporal and non spatial reality'. This underlying reality is however structured in such away that `if we move through it, this influences the way time flows in case we decide to coordinate our movement in a coordinate system'.

Also with respect to the Minkowski structure of time-space and the $\real^3$ structure of space, we have to inverse things. First and much more primitive is the hyperbolic structure of Minkowski. It is only much afterwards -- most probably influenced profoundly by our local situation of very small sizes and velocities, and we want to analyse in great detail how this influence took shape -- that we identified a three dimensional space structure as if it could be separated out. This is not the case, and due to an error of believing that `places can be looked at independently of these places also evolving in time'. `Time always flows for us, but also for every material entity, and that is the reason that places -- i.e. space -- can not be separated off without the danger of errors, we want to identify very exactly these errors. By the way, we believe that a lot of confusion and not understanding of relativity is due to the belief that it is linked to `observers'. We do not believe so. Relativity describes the reality of the material entities, also when no observations at all are present. Since our bodies are material entities we are also with our bodies part of this relativistic reality.

\section{The kinematic reality beneath space-time}
Following the traditional approach it is not possible to reflect properly in the Minowski framework, it never is clearly stated and/or explained what the meaning is of all the four vectors that are introduced to built the whole theory, its kinematic and its dynamics. Let us see whether our approach can shed light on these aspects. For example, what is the four velocity?

The four velocity $U=(U_0, U_1, U_2, U_3)$ is the derivative with respect to the intrinsic time $\tau$ of the four components of the coordination in time-space. Let us make some calculation to see in which way the four velocity is connected with the three velocity.
We have
\begin{eqnarray}
U_0={dX_0 \over d\tau}={dX_0 \over dt}{dt \over d\tau}=c{dt \over d\tau}
\end{eqnarray}
And we have
\begin{eqnarray}
(cd\tau)^2=(cdt)^2-(dx_1)^2-(dx_2)^2-(dx_3)^2 \\
\Rightarrow ({cd\tau \over cdt})^2=1-({dx_1 \over cdt})^2-({dx_2 \over cdt})^2-({dx_3 \over cdt})^2 \\
\Rightarrow ({d\tau \over dt})^2=1-{v^2 \over c^2} \\
\Rightarrow {d\tau \over dt}=\sqrt{1-{v^2 \over c^2}} \\
\Rightarrow {dt \over d\tau}={1 \over \sqrt{1-{v^2 \over c^2}}}
\end{eqnarray}
From this follows that the `time component' $U_0$ of the four velocity is given by 
\begin{eqnarray}
U_0={1 \over \sqrt{1-{v^2 \over c^2}}}c
\end{eqnarray}
For the space components we have 
 \begin{eqnarray}
 U_i={dX_i \over d\tau}={dX_i \over dt}{dt \over d\tau}={1 \over \sqrt{1-{v^2 \over c^2}}}v_i
\end{eqnarray}
In our case, were we only consider one time axis and one space axis, it is given by 
 \begin{eqnarray}
 U={1 \over \sqrt{1-{v^2 \over c^2}}}(c,v)
 \end{eqnarray}
 Note that the size of this vector is given by 
 \begin{eqnarray}
 \|U\|=\sqrt{\langle U|U\rangle}=\sqrt{c^2-v^2 \over {1-{v^2 \over c^2}}}=c
 \end{eqnarray}
 Well, well, this is of course the surf velocity. So ``the velocity of light in physical reality is the surf velocity in our example of the World-Wide Web". Of course, we need to keep taking into account the counterintuitive nature of the Minkowski metric. Although the surf velocity is always equal to the velocity of light, the time and space components of this velocity can be very different. In fact, for a physical entity moving with constant velocity $v$, the space component   is ${1 \over \sqrt{1-{v^2 \over c^2}}}v$, which is a magnitude between $v$ and infinity, and increasing towards infinity with increasing value of $v$. This expresses that indeed for the moving entity speed goes up, equally so as time dilates and distance contracts. The time component behaves similarly with respect to the velocity $c$, which means that its magnitude is between $c$ and infinity, and increases with increasing velocity. In fact, with respect to the intrinsic experience of the moving physical entity $B$, $c$ is obviously a limit velocity to be obtained as measured by the not moving physical entity $A$. It means `infinite velocity with respect to the time coordinate of $A$ and infinite velocity with respect to the space coordinate of $A$'. If we go back to our example of surfing on the World-Wide Web, we can understand it, when starting from $A_0=B_0$ it is possible to jump without clicking any links right away to $A_m=B_l$, then in the scheme where links are counted this means `infinite velocity' in all possible dimensions, be it the time or the space one. Note that the `possibly bigger than $c$ velocity' intrinsic velocity in the space realm of $B$' becomes `smaller than $c$ not intrinsic velocity in the space realm of $A$', because there is `time dilatation' and `length contraction' in the $A$ frame, which makes hence `appear' the intrinsic velocity possibly bigger than $c$ for $B$ as a not intrinsic velocity smaller than $c$ for $A$. Hence, if we -- being in the $A$ frame -- see something -- the $B$ entity -- move with velocity $v$ in the space realm, i.e. measuring the distance run per unit time in our $A$ frame, then this $B$ entity is actually moving with an intrinsic velocity ${1 \over \sqrt{1-{v^2 \over c^2}}}v$ bigger than $v$ and possibly also bigger than $c$ in the space realm. We can even easily calculate when this intrinsic velocity bigger than $v$ also becomes bigger than $c$. This happens when
\begin{eqnarray}
 {1 \over \sqrt{1-{v^2 \over c^2}}}v=c \\
 \Leftrightarrow 
v= \sqrt{1-{v^2 \over c^2}}c \\
 \Leftrightarrow 
v^2= c^2-v^2 \\
 \Leftrightarrow 
v= {c \over \sqrt{2}}=0.7071 c
\end{eqnarray}
This shows that what concerns the intrinsic velocity of a physical entity in the space realm, the velocity of light is not a limit, and even not a special value to pass by. The velocity of light $c$ only plays this role of `maximum to be reached velocity' if the space realm of a physical entity -- the entity $B$ in our case -- is looked at from the space realm of another physical entity -- the entity $A$ in our case.

Let us, by way of example, calculate these time and space coordinate for the example in Figure 1. We have for this situation that 
\begin{eqnarray} \label{velocityAB}
v=c\sqrt{{m^2-l^2 \over m^2}}=c\sqrt{10^2-8^2 \over 10^2}=0.6 c=0.6
\end{eqnarray}
if we put $c=1$ like in Figure 1. This means that 
\begin{eqnarray}
{1 \over \sqrt{1-{v^2 \over c^2}}}={1 \over \sqrt{0.64}}={1 \over 0.8}={5 \over 4}
\end{eqnarray}
This means that the $x$ component of the four velocity of $B$ equals 
\begin{eqnarray}
U_1=0.6 \cdot {5 \over 4}={3 \over 4}
\end{eqnarray}
which is $75\%$ of the velocity of light, while $v$, the velocity of $B$ measured in the time-space frame of $A$ only values $60\%$ of the velocity of light.

The time component of the four velocity is 
\begin{eqnarray}
U_0={5 \over 4}
\end{eqnarray}
which is $125\%$ of the velocity of light. And indeed, $B$ moves more quickly from $A_0=B_0$ to $A_m=B_l$, needing 8 years, than $A$ moves from $A_0=B_0$ to $A_m=B_l$, needing 10 years.

Let us calculate the size of the four vector. We have 
\begin{eqnarray}
U_0^2-U_1^2=25/16-9/16=16/16=1
\end{eqnarray}
Hence 
\begin{eqnarray}
\|U\|=\sqrt{\langle U|U\rangle}=\sqrt{1}=1
\end{eqnarray}
which shows that intrinsically $B$, in its own reference frame, moves with the velocity of light, like does $A$ in its own reference frame. This shows well that we can interpret the velocity of light as being the `surfing velocity'. Indeed, also in our example of the World-Wide Web, we suppose that $A$ and $B$ surf with the same velocity, but $A$ takes 10 click to arrive at the meeting point while $B$ only takes 8 clicks.
 
This is in fact the moment to stand still somewhat longer with the counter intuitive aspects of the Minkowski metric, in an attempt to identify more deeply its root. Indeed, if we imagine $A$ and $B$ surfing with the same surf velocity $c$, and starting together at $A_0=B_0$, then our intuition tells us that they will not arrive at the same time at $A_m=B_l$, namely $B$ will arrive way before $A$ arrives there. To arrive together and meet there $A$ would have to surf with a higher speed. The reason for our intuition telling us this is because also for the example of the World-Wide Web we imagine it happening in a Newtonian time-space. We refer, for example, to a watch that $A$ and $B$ would carry along, and we want to identify the meeting place $A_m=B_l$ only as a `spot in space' not linked to time. Of course this ``is" how real life surfing would take place. Perhaps we would come closer to imagining an analogy if we take distance from the World-Wide Web, and think about two persons $A$ and $B$, meeting in a pub $A_0=B_0$, having a specific discussion about some `meaning issue', and the meeting again later $A_m=B_l$ in a pub, were they discuss again. In between $A$ has made 10 conceptual steps while $B$ has made 8 conceptual steps. Again one could, if referring to the Newtonian world, say that $B$ reflected faster than $A$. Another way is to see it as if $B$ was able to take a shorter conceptual path as compared to $A$ while both were all the time reflecting at the same speed. Minkowski metric and its reality teaches us that physical entities behave in the second way. Our difficulty of being able to grasp this clearly is due to our earthly bounded experience with reality, were we can mainly identify `spots' in space as being the reference for `meetings'. That this does not bring our bodies -- which are physical entities -- in deep trouble, is due to our bodies being very small and almost always being at rest with respect to other physical entities. So, we have to make a specific effort with our mind to imagine the Minkowski metric related reality, and this is not a question of `observers related matters'. It is how realty is, and we better try to take it seriously.


Let us go back now to the physical realm. The above insight means that material physical entities `move' through the non temporal and non spatial realm of reality with a constant velocity $c$, which is equal to the velocity of light. Let us come to light itself now. Light is not a material physical entity, it has no mass, it is a boson, and not a fermion, etc... This means that the way of interacting of light with material physical entities is different from surfing. There are no clicks moving from one page to another involved.

The behavior of light is a singularity of the mechanical equations of relativity. These mechanical equations are principally about the behavior of physical entities with mass, and the massless entities, photons, are at the edge of this mechanical theory. We will see that a statement such as `light has velocity $c$ when moving through the vacuum of the space that we constructed to give place to all physical entities', is more subtle than imagined. 

That the `surfing velocity' is independent of the reference frame is natural, because it is a non temporal and non spatial velocity. Does this mean that there is an aether involved? No, because the notion of aether is a notion that only makes sense when time and space are already created and physical entities immersed in its coordination. One could say that the surfing velocity being independent of the reference frame means that there is an underlying non temporal and non spatial reality, it is within this reality that this surfing velocity exists. It only reveals itself following our analysis also when a time-space reference frame is attempted to coordinate world lines. And at first place it reveals itself as the magnitude of the four velocity. Only at second place it also reveals itself as `how photons which do not have mass move through space to connect webpage without clicking'. Indeed, what is the intrinsic velocity in the space realm of a photon? To be able to answer this question, we need to know the velocity four vector of a photon. In text books on relativity theory it is generally stated that such a velocity four vector is not well defined, and hence that it only exists for massive physical entities. However, we can in a meaningful way consider the photon as a limit particle, and hence also calculate its velocity four vector as this limit. We find then $\infty$ for the time component and also $\infty$ for each of the space components. But its size is equal to $c$ as is the case for a massive physical entity. A photon travels with surf velocity equal to $c$ in the non temporal and non spatial realm, but in its own time-space frame it does not need a click to move in the time direction, nor needs to run through length units in its space realm. This is the Minkowsky nature of physical reality in it weirdest aspect. Indeed, when looked at the photon from any other frame, it moves exactly with velocity $c$ in the time-space of this other frame. It is the `any other' which is at the origin of the constancy of this velocity of the photon in any reference frame. But, let us remember that also this velocity $c$, exactly like any velocity $v$ of any massive physical entity, is `not' an intrinsic property of the entity in question. It is a property changed due to it being looked at from another reference. This is equally so for a photon, the velocity $c$ is `not' its intrinsic space realm velocity, because this one is infinite.

\section{The dynamic reality beneath space-time}

What about relativistic dynamics? Often it is said that when the velocity $v$ increases the mass increases and that this is the reason why a physical entity cannot reach the velocity of light, its mass going to infinity with increasing velocity. We believe that this is a wrong way of looking at dynamical aspects of the situation. Indeed, remember that the space component of the four velocity is given by 
\begin{eqnarray}
U={1 \over \sqrt{1-{v^2 \over c^2}}}v
\end{eqnarray}
which means that this component goes to infinity when $v$ goes to $c$. Of course, also the time component of the four velocity being equal to 
\begin{eqnarray}
U_0={1 \over \sqrt{1-{v^2 \over c^2}}}c
\end{eqnarray}
goes to infinity when $v$ goes to $c$. That the magnitude of the velocity remains finite, and constant equal to $c$, is linked to the special form of the Minkovski metric, which subtracts two infinities, which of course can lead to a finite quantity. All this means that already on the level of the kinematics the behavior of light -- moving through space with velocity $c$ -- is singular. Mass however does not behave singularly at all in our opinion, rather on the contrary, it is an invariant, hence an intrinsic property of the physical entity under consideration. Of course, this is well known, and acknowledged in standard textbook relativity, for what concerns `rest mass'. We believe that it is the only mass that exists, the one that is measured well when a physical entity is at rest with respect to the reference frame were we measure its mass. The apparent increase of mass with velocity is due to the increase of the spatial coordinate of the four velocity with velocity. Hence, we believe that one should not speak about `relativistic mass', like in many textbooks on relativity theory. The four momentum $P$ of a massive physical entity is the mass multiplied by the four velocity $U$, hence 
\begin{eqnarray}
P=mU
\end{eqnarray}
Let us note right away that we can calculate the magnitude of the four momentum, and then get 
\begin{eqnarray}
\|P\|=\sqrt{\langle P|P\rangle} =m\sqrt{\langle U|U\rangle}=mc
\end{eqnarray}
The size of the momentum of a massive physical entity is also an invariant, and it is the momentum of a mass $m$ with velocity $c$. In our example, it is the surf momentum. What could in our example be the equivalent of mass? We think of `meaning impact', i.e. the size with which the meaning impacts -- we use even in every day language the expression `impact' when it concerns meaning. Of course, this again only makes sense in case we see surfing not just as a passive action with respect to fixed webpages, but as a dynamical action, were every visit of a webpage also potentially changes the meaning content of this webpage. When two massive physical entities collide, the equivalent would be that two webpages come into competition, both wanting to occupy the same state. If both are solids -- webpages that are very stubborn in adapting and/or making compromise in meaning content with each other -- the collision can be of the elastic type. But collisions can also lead to merging giving rise to a third webpage containing a consensus of the two colliding ones.

If we calculate the product of the four momentum with itself, and express it to be equal to the square of the invariant which is the size of the momentum, we find the famous formula of relativity, expressing energy in function of mass. That, of course, is also linked to interpreting the time component of the four momentum as `the energy divided by the velocity of light', hence 
\begin{eqnarray}
P=({E \over c}, p_1, p_2, p_3)
\end{eqnarray}
Let us make the calculation. We have 
\begin{eqnarray}
\langle P|P\rangle={E^2 \over c^2}-p_1^2-p_2^2-p_3^2=m^2c^2
\end{eqnarray}
If we consider the situation of a physical entity at rest, which means that $p_1=p_2=p_3=0$, we get 
\begin{eqnarray}
E=m c^2
\end{eqnarray}
which is the famous formula derived by Albert Einstein in his 1905 work on the theory of special relativity \cite{einstein1905b}. 
Let us remark that a part of the energy comes from the momentum $mc$ that any mass carries with itself in its `flow in time' -- and taking into account the existence of the non temporal underlying reality, we should say `the momentum that the mass, i.e the meaning impact, carries while moving through the non temporal reality, moving through the overall meaning structure, and that this appears like `moving in time' is due to time-space coordination creation of the situation. So, instead of interpreting mass-energy as potential energy, we can now interpret it as kinetic energy. If mass is turned into light energy, this is as `taking away  the huge momentum that the mass has in its flow through time -- its motion of surfing' and giving it away in the form of light.

What about light? Since in standard relativity textbooks one considers that no meaning can be given to the four velocity for light, also the formula $P=mU$ is not considered valid for light. However, in our case, were we have considered $U$ to exist for light, we can investigate whether we can keep $P=mU$ as valid for light too? Let us first consider the value of the four momentum.
For $m=0$ we have 
\begin{eqnarray}
\langle P|P\rangle={E^2 \over c^2}-p_1^2-p_2^2-p_3^2=0
\end{eqnarray}
which leads to 
\begin{eqnarray}
E=cp
\end{eqnarray}
where $p$ is the magnitude of the three momentum, and hence
\begin{eqnarray}
P=(p, p_1, p_2, p_3)
\end{eqnarray}
which shows that $P$ is a null four vector, located on the light cone, which we would expect for light. The sizes of $p_1$, $p_2$, $p_3$ and $p$ are determined by quantum theory. 

Let us remark explicitly that Figure 1, since it is drawn in a reference frame $t, x$, might give the impression that the points of this place are points of a Euclidean plane. This is another -- although related to the one we mentioned before -- aspect what makes the Minkowski metric so counter intuitive. The path taken by $B$, if we look at the $t, x$ frame as an Euclidean plane obviously is much longer than the path taken by $A$. While following the Minkowski metric it is the other way around. Light which takes a path inclined $45^\circ$ is the shortest possible, while in an Euclidean view of the plane used in Figure 1 this is even longer than both paths, the one taken by $A$ and the one taken by $B$. Our minds, when they see a plane, only can imagine this plane to be Euclidean. We can imagine a curvature that deviates in only a smooth way from Euclidean, since we know what twisted pieces of paper are, for example. But Minkowski really deviates profoundly from Euclidean, so our imagination fails about it. However, experiments confirming abundantly the theory of relativity show us that `this is the way reality is'. When space is explored starting from time and an underlying non temporal and non spatial reality, in the way we analyzed above, then the plane depicting a time and a space coordinate is Minkowskian and not Euclidean. This is `not' due to observation, hence the traditional way that relativity is attempted to be explained obscures its essence in our opinion. The Minkowski structure arises from deep reality itself, and is explored by physical entities with mass and photons of light. Human beings and more even human observers do not play any role in it. However, in the fact that we find the Minkowski metric so counter intuitive when we see it exposed, and more even when we see it drawn on a plane, the reason for this is deeply rooted in the nature of human observation. We are used to see paths in space only without involving the effect that velocity has on reaching into the future. The reason is that we are small, surrounded by all small physical entities, and we and all customary physical entities around us move `in space' -- remember that every physical entity moves with the velocity of light in time-space, or, more correctly in the non temporal and non spatial underlying reality -- very very slowly only.

Light does play a crucial role in our experience too. But light, because it consists of physical entities with zero mass, cannot be `joined by us'. We, bound to our physical body which has mass different from zero, when we want to introduce a time-space frame for ourselves, `cannot do this joining the time-space frame were light is in'. This means that we can only observe light from the outside, in our time-space frame, and not in its own. That is the reason that light shows itself to us in the way we perceive it, namely `always' moving with a space -- in our time-space frame -- velocity equal to $c$, exactly the same quantity with which we move in our own time-space frame in the direction of our four velocity vector, which means, partly in our time direction, and partly in our space direction -- were also we see light moving. 

Light moves so quickly `in the space realm of any time-space frame from were we observe it', namely with speed $c$ that we do not see it moving in our space realm, so we are only confronted with other properties of light than the ones that could us make aware of Minkowski nature of reality directly. Indirectly it is light that showed us the way, first in its appearance in the form of electricity and magnetism, leading to Maxwell's equations. Light might also carry some of the other keys to the deeper reason for the exact structure of space, i.e. Euclidean, since it is so much present in it.

Can we understand more profoundly how our Euclidean intuition has grown and misguided us in our image about time-space? We indeed can. Space as a distinct entity with three dimensions arose as a consequence of further exploration of the time-space in the way we analyzed above. Till now we have neglected to make a difference between a situation were there would be more than one dimension of `space' and a situation with only one space dimension. But, the Minkowski metric of the experienced mechanics and electromagnetics of our world gives rise to a three dimensional space. It is, in fact, also this Euclidian nature of three dimensional space which makes it too difficult for us to intuitively imagine the nature of hyperbolic time-space. How are we confronted with this more than one dimension of space and its Euclidean nature?  Let us investigate this question.

\section{The reality of three dimensional space}

If we consider only two world lines $(A_n)_n$ and $(B_k)_k$ elements of ${\cal A}_{0,m}$, like we did so far, we will not need more than one space dimension to fit them in the Minkowski structure. Hence, let us consider a third world line $(C_i)_i$ of a physical entity $C$ element of ${\cal A}_{0,m}$. This means that we have $A_0=B_0=C_0$ and $A_m=B_l=C_j$. We necessarily also have $j \le m$. With respect to the values of $l$ and $j$ different possibilities exist. Let us analyse what happens with the different velocities that come into play as a consequence of the Minkowski metric. Without loss of generality we can suppose that $j \le l$, which means that $(C_i)_i$ is the fastest of the three considered paths in heading towards the reunion of the three. The velocity $v^{A}_{C}$ with which $C$ moves in the time-space coordination of $A$ is given by 
\begin{eqnarray}
v^{A}_{C}=c\sqrt{m^2-j^2 \over m^2} 
\end{eqnarray}
which is bigger than the velocity $v^{A}_{B}$ with which $B$ moves in the time-space coordination of $A$
\begin{eqnarray}
v^{A}_{B}=c\sqrt{m^2-l^2 \over m^2}
\end{eqnarray}
because $j \le l$. 
The first question we want to consider is `whether the nature of the Minkowski co-ordination of our non temporal and non spatial reality' indicates the existence of at least such a $C$, different from $A$ and $B$. The answer is `yes', but more, we will also consider the situation where $j=l$, and hence $B$ and $C$ are, what concerns the time needed to reach the meeting point $A_m=B_l=C_j$ the two paths $B$ and $C$ are the same. And even in this more simple situation we will see that the Minkowski structure induces more than one space dimension. Let us consider for a moment the situation of Figure 1, but now with $B$ and $C$ moving in opposite direction. Then, already for this situation, Minkowski allows for $(C_i)_i$ to be a different path than $(B_k)_k$, although both paths are located in one and the same space direction in this case, which means that by means of this case no extra space dimension is yet induced by the Minkowski structure. For this we will have to consider a situation were $(C_i)_i$ moves in another direction, not just the opposite of the direction in which $(B_k)_k$ moves.

We start our investigation by considering the Lorentz transformation that connects the coordinates of a four vector in the $B$ time-space coordination with the coordinates of the four vector in the $A$ time-space coordination. This Lorentz transformation is given by
\begin{eqnarray}
L(B,A)=\left( \begin{array}{cccc}
{1 \over \sqrt{1-({v^{A}_{B} \over c})^2}} & {-{v^{A}_{B} \over c} \over \sqrt{1-({v^{A}_{B} \over c})^2}} & 0 & 0  \\
{-{v^{A}_{B} \over c} \over \sqrt{1-({v^{A}_{B} \over c})^2}} & {1 \over \sqrt{1-({v^{A}_{B} \over c})^2}} & 0 & 0 \\
0 & 0 & 1 & 0 \\
0 & 0 & 0 & 1  \end{array} \right) 
\end{eqnarray}
Let us now consider the situation of Figure 1, hence for the $(A_n)_n$ world line we have $m=10$ and for the $(B_k)_k$ world line we have $l=8$. We consider $(C_i)_i$ with $j=l=8$. We make the hypothesis that for $B$ a space dimension $x$ comes into being coordinating the more speedy motion of $B$ towards the meeting point with $A$ at $A_{10}=B_{8}$. And now we add the hypothesis that for $C$ another space dimension $y$ comes into being coordinating the speedy motion of $C$ towards the meeting point at $A_{10}=B_{8}=C_{8}$. We have represented this situation in Figure 2.
\begin{figure}[htbp]
\begin{center}
\includegraphics[scale =1]{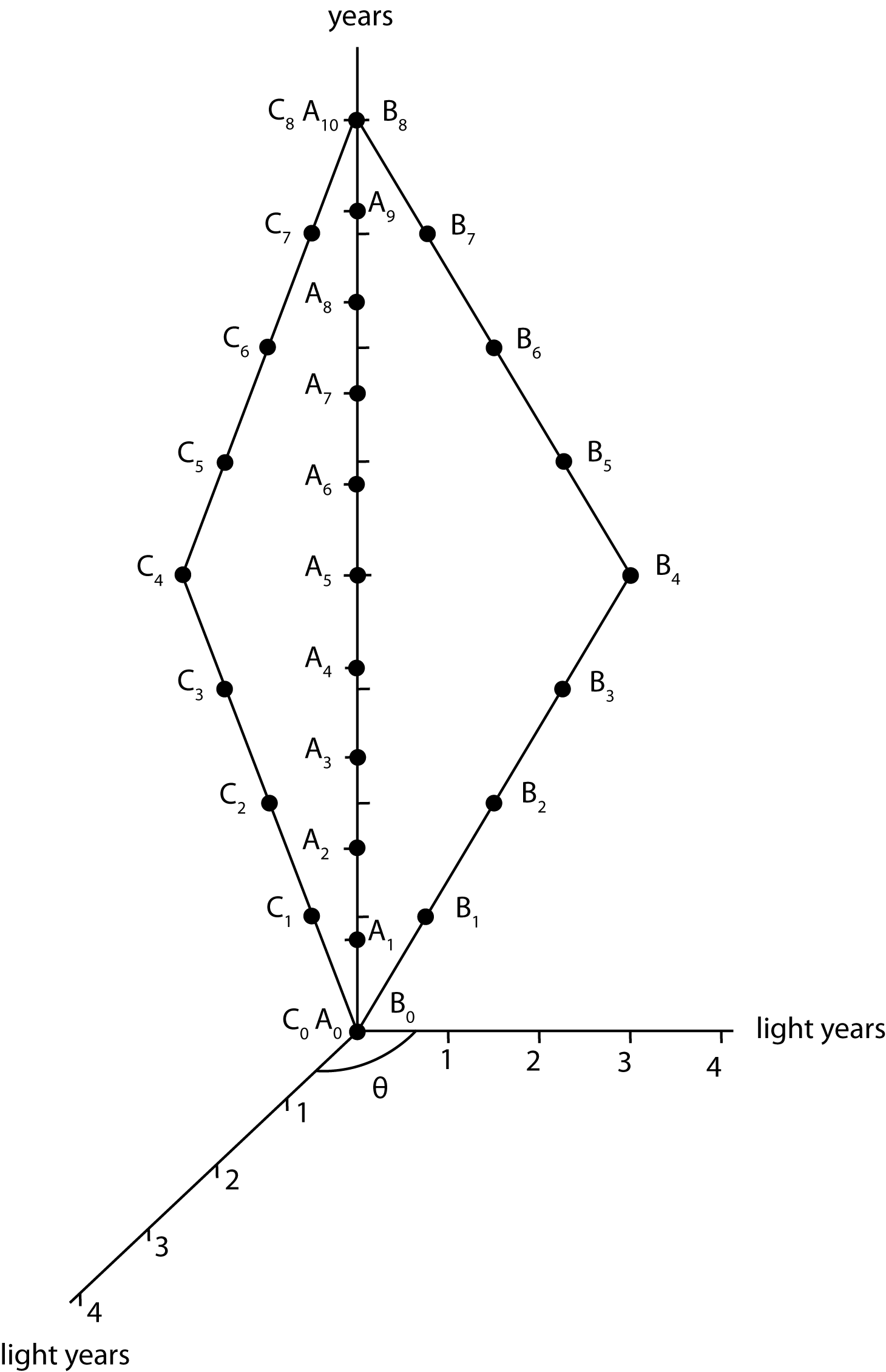}
\caption{A graphical representation of $A$, $B$ and $C$ for $m=10$ and $l=j=8$}
\end{center}
\end{figure}
Let us write down the four vectors of velocity for $A$, $B$ and $C$ in the $A$ time-space frame that expresses the above hypothesis. We have
\begin{eqnarray}
U^{A}_{A}={1 \over \sqrt{1-({v^{A}_{B} \over c})^2}}(c, 0, 0, 0) \\
U^{A}_{B}={1 \over \sqrt{1-({v^{A}_{B} \over c})^2}}(c, v^{A}_{B}, 0, 0) \\
U^{A}_{C}={1 \over \sqrt{1-({v^{A}_{C} \over c})^2}}(c, v^{A}_{C}\cos\theta, v^{A}_{C}\sin\theta, 0)
\end{eqnarray}
where $v^{A}_{B}$ and $v^{A}_{C}$ are the sizes of the `space velocities' related to $B$ and to $C$, and $\theta$ is the angle which parametrizes the direction of the space velocity delated to $C$ in the $xy$ plane. So, for the situation that we consider in Figure 1, and added $C$, we have
\begin{eqnarray}
v^{A}_{B}=v^{A}_{C}=0.6c \quad \theta \in [0, 2\pi]
\end{eqnarray}
This gives us
\begin{eqnarray}
U^{A}_{A}={5 \over4}(c, 0, 0, 0) \\
U^{A}_{B}={5 \over4}(c, {3 \over 5}, 0, 0) \\
U^{A}_{C}={5 \over4}(c, {3 \over 5}\cos\theta, {3 \over 5}\sin\theta, 0)
\end{eqnarray}
\begin{eqnarray}
L(B,A)=\left( \begin{array}{cccc}
{5 \over 4} & -{3 \over 4} & 0 & 0  \\
-{3 \over 4} & {5 \over 4} & 0 & 0 \\
0 & 0 & 1 & 0 \\
0 & 0 & 0 & 1  \end{array} \right) 
\end{eqnarray}
We can calculate the four vectors in the $B$ time-space frame. This gives
\begin{eqnarray}
U^{B}_{A}=L(B,A)U^{A}_{A} \\
U^{B}_{B}=L(B,A)U^{A}_{B} \\
U^{B}_{C}=L(B,A)U^{A}_{C}
\end{eqnarray}
and hence
\begin{eqnarray}
U^{B}_{A}=\left( \begin{array}{c}
U^{B}_{A0}  \\
U^{B}_{A1} \\
U^{B}_{A2}  \\
U^{B}_{A3} \end{array} \right)=\left( \begin{array}{cccc}
{5 \over 4} & -{3 \over 4} & 0 & 0  \\
-{3 \over 4} & {5 \over 4} & 0 & 0 \\
0 & 0 & 1 & 0 \\
0 & 0 & 0 & 1  \end{array} \right) \left( \begin{array}{c}
{5 \over 4} c  \\
0 \\
0  \\
0  \end{array} \right) = \left( \begin{array}{c}
{25 \over 16} c  \\
 -{15 \over 16} c \\
0  \\
0  \end{array} \right) \\
U^{B}_{B}=\left( \begin{array}{c}
U^{B}_{B0}  \\
U^{B}_{B1} \\
U^{B}_{B2}  \\
U^{B}_{B3} \end{array} \right)=\left( \begin{array}{cccc}
{5 \over 4} & -{3 \over 4} & 0 & 0  \\
-{3 \over 4} & {5 \over 4} & 0 & 0 \\
0 & 0 & 1 & 0 \\
0 & 0 & 0 & 1  \end{array} \right) \left( \begin{array}{c}
{5 \over 4} c  \\
{3 \over 4} c \\
0  \\
0  \end{array} \right) = \left( \begin{array}{c}
{25 \over 16} c-{9 \over 16} c  \\
 -{15 \over 16} c + {15 \over 16} c \\
0  \\
0  \end{array} \right) \\
U^{B}_{C}=\left( \begin{array}{c}
U^{B}_{C0}  \\
U^{B}_{C1} \\
U^{B}_{C2}  \\
U^{B}_{C3} \end{array} \right)=\left( \begin{array}{cccc}
{5 \over 4} & -{3 \over 4} & 0 & 0  \\
-{3 \over 4} & {5 \over 4} & 0 & 0 \\
0 & 0 & 1 & 0 \\
0 & 0 & 0 & 1  \end{array} \right) \left( \begin{array}{c}
{5 \over 4} c  \\
{3 \over 4} c \cos\theta \\
{3 \over 4} c \sin\theta  \\
0  \end{array} \right) = \left( \begin{array}{c}
{25 \over 16} c-{9 \over 16} c \cos\theta  \\
 -{15 \over 16} c + {15 \over 16} c \cos\theta \\
{3 \over 4} c \sin\theta  \\
0  \end{array} \right)  
\end{eqnarray}
Hence we find
\begin{eqnarray}
U^{B}_{A}={5 \over4}({5 \over 4}c, -{3 \over 4}c, 0, 0) \\
U^{B}_{B}=(c, 0, 0, 0) \\
U^{B}_{C}=(c+{9 \over 16} c(1- \cos\theta) , -{15 \over 16} c (1 - \cos\theta), {3 \over 4} c \sin\theta, 0)
\end{eqnarray}
which makes it possible to calculate the space velocities in the $B$ time-space frame. We find
\begin{eqnarray}
v^{B}_{A}={U^{B}_{A1} \over U^{B}_{A0}}c=(-{3 \over 5}c, 0, 0)=(-0.6c, 0, 0)
\end{eqnarray}
This is what we expect. If $B$ moves with space velocity $0.6c$ in the positive $x$-direction in the $A$ time-space frame, then $A$ moves in the $B$ time-space frame with the same velocity in the opposite direction, hence the negative $x$-direction. We also find
\begin{eqnarray}
v^{B}_{B}={U^{B}_{B1} \over U^{B}_{B0}}c=(0, 0, 0)
\end{eqnarray}
This is also what we expect. In the $B$ time-space frame $B$ is at rest. And we also find
\begin{eqnarray}
v^{B}_{C}&=&{U^{B}_{C1} \over U^{B}_{C0}}c=({-{15 \over 16} c (1 - \cos\theta) \over c+{9 \over 16} c(1- \cos\theta)}, {{3 \over 4} c \sin\theta \over c+{9 \over 16} c(1- \cos\theta)}, 0)c=(-{{15 \over 8}c\sin^2{\theta \over 2} \over 1+{9 \over 8}\sin^2{\theta \over 2}}, {{3 \over 2}c\sin{\theta \over 2}\cos{\theta \over 2} \over 1+{9 \over 8}\sin^2{\theta \over 2}}, 0) \\
&=&{3 \over 2}c\sin{\theta \over 2}(-{{5 \over 4}\sin{\theta \over 2} \over 1+{9 \over 8}\sin^2{\theta \over 2}},{\cos{\theta \over 2} \over 1+{9 \over 8}\sin^2{\theta \over 2}}, 0)
\end{eqnarray}
What is now very interesting is that the magnitude of $v^{B}_{C}$, hence $|v^{B}_{C}|$ is no longer constant, and changes with $\theta$. This means that we can `detect' the presence of a second dimension by purely measuring the magnitude of the space velocity in the $B$ time-space frame. The reason is very deep, namely that the Lorentz transformation only provokes a length contraction in the direction of the space velocity of the moving reference frame, and not in the other directions.
Let us calculate the size $|v^{B}_{C}|$ of $v^{B}_{C}$. We get
\begin{eqnarray}
|v^{B}_{C}(\theta)|=\sqrt{(v^{B}_{C1})^2+(v^{B}_{C12})^2}={3 \over 2}c\sin{\theta \over 2}\sqrt{({{5 \over 4}\sin{\theta \over 2} \over 1+{9 \over 8}\sin^2{\theta \over 2}})^2+({\cos{\theta \over 2} \over 1+{9 \over 8}\sin^2{\theta \over 2}})^2}
\end{eqnarray}
Let us calculate some specific values. We get
\begin{eqnarray}
|v^{B}_{C}(0)|=0 \\
|v^{B}_{C}(\pi)|={3 \over 2}c({{5 \over 4} \over 1+{9 \over 8}})={3 \over 2}c({{10 \over 8} \over {17 \over 8}})={3 \over 2}c({10 \over 17})={15 \over 17}c=0.88c \\
|v^{B}_{C}({\pi \over 2})|={3 \over 2}c{\sqrt{2} \over 2}\sqrt{({{5 \over 4}{\sqrt{2} \over 2} \over 1+{9 \over 8}{1 \over 2}})^2+({{\sqrt{2} \over 2} \over 1+{9 \over 8}{1 \over 2}})^2}={3\sqrt{2} \over 4}c\sqrt{({{5\sqrt{2} \over 8} \over {25 \over 16}})^2+({{\sqrt{2} \over 2} \over {25 \over 16}})^2} \\
={3\sqrt{2} \over 4}c\sqrt{({10\sqrt{2} \over 25})^2+({8\sqrt{2} \over 25})^2}={3\sqrt{2} \over 4}c\sqrt{{200 \over 625}+{128 \over 625}}={3 \over 4}\sqrt{{656 \over 625}}c=0,77c
\end{eqnarray}
Hence, for a velocity of $v^{A}_{C}$ equal in size, namely $0.6c$, and in the same direction as $v^{A}_{B}$ in the $A$ time-space frame, which is what the value $\theta = 0$ represents, we get a velocity $|v^{B}_{C}(0)|=0$ in the $B$ time-space frame. This is what we would expect, and we considered already this case above.
For a velocity of $v^{A}_{C}$ equal in size, namely $0.6c$, but in opposite direction to $v^{A}_{B}$ in the $A$ time-space frame, which is what the value $\theta = \pi$ represents, we get a velocity $|v^{B}_{C}(0)|=0.88c$. This situation would still not need an extra space dimension, and it also correspond to what we might expect from the well-known sum-of-relativistic-velocity formula. Here comes an interesting new type case. For a velocity of $v^{A}_{C}$ equal in size, namely $0.6c$, but in a direction orthogonal to $v^{A}_{B}$ in the $A$ time-space frame, which is what the value $\theta = {\pi \over 2}$ represents, we get a velocity $|v^{B}_{C}(0)|=0.77c$. This `size of space velocity' in the $B$ time-space frame cannot be explained without the introduction of more than one space dimension. In Figure 3 we have represented the graph of the sizes of the space velocity $|v^{B}_{C}(\theta)|$ for different values of $\theta$. The scale of the $\theta$ variable is divided in 20, which means $18^\circ$ for each unit.
\begin{figure}[htbp]
\begin{center}
\includegraphics[scale =0.75]{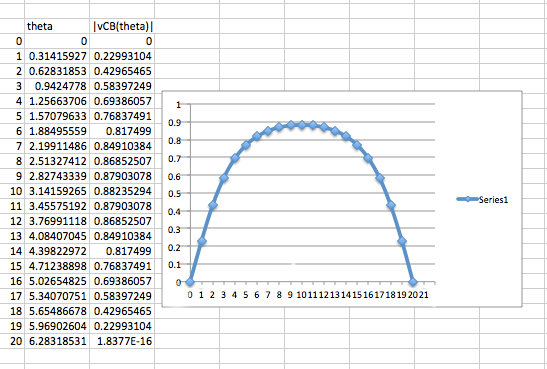}
\caption{A graphical representation of $|v^{B}_{C}(\theta)|$ for different values of $\theta$}
\end{center}
\end{figure}
We can see that $|v^{B}_{C}(\theta)|$ increases from $0$ to ${15 \over 17}c=0.88c$ when $\theta$ varies from $0$ to $\pi$, and then decreases from ${15 \over 17}c=0.88c$ to $0$ when $\theta$ varies from $\pi$ to $2\pi$. All the intermediate values can be measured in the $B$ time-space frame `as sizes of the space velocity of $C$'. This is an straight forward indication of the `necessity to introduce more than one space dimension' if coping with the `facts of reality and the data in sizes of space velocities extracted from it' that we can move with different four velocities towards meeting points. Or, if we go back to our surf and World-Wide Web example, we need more than one space dimension if we want to cope with `space and space velocities' as remedies for the fact that we can reach a meeting point by different number of clicks, and in such a way that (i) there is a `most slow path', i.e. a path needing the most clicks, and (ii) there are multitudes of faster paths that are different.

\section{Quantum and relativity}

We have mentioned that our analysis is inspired by what we have learned in the context of quantum theory. However, so far we have only talked about material entities with mass and about light, without giving any thought to how quantum entities would behave in terms of relativity. Many approaches have been worked out to find a way to reconcile the principles of quantum theory with those of relativity, but without obtaining any truly clear view of the matter, and the less so if attempts are made to include gravity as well. In this section, we will give only an outline of ideas and of their aspects that we want to investigate further.

The greatest success story of quantum and relativity is that of the Dirac equation for the electron \cite{dirac1928}. With exceptional ingenuity, Dirac constructed a wave equation of the first order in time and space -- i.e. handling time and space on the same footing, as opposed to  Schr\"odinger's equation, which is first order in time and second order in space -- leading to a `squared' version compatible with the well-known second order in time and second order in space `wave equations' of classical electrodynamics. In this way, the spin of the electron spontaneously made its appearance, additionally introducing two new components, which led to the theoretical description of what turned out to be anti-matter. It is known that anti-matter can also be looked upon as matter `moving in the opposite direction in time'. We want to investigate whether our new view on relativity theory can provide us with a possible understanding of these issues.

If matter `moves' at a velocity $c$ surfing over a non-spatial and non-temporal reality in a specific direction, it is not difficult to imagine that it will also be possible for matter to surf in a direction exactly opposite to that specific direction. But matter which surfs in the opposite direction will reveal itself as anti-matter when looked upon from a reference frame connected with matter surfing in the specific direction.
The same connection of time-space creation with a surfing movement above a non-temporal and non-spatial reality can explain the time-inversion symmetry of electromagnetic laws.

What about spin appearing spontaneously when Dirac constructed a first order in time and space wave equation? In our opinion, but of course this needs to be studied in depth, the phenomenon should be looked upon inversely. Namely, time-space as a creation upon the surfing movement can only give rise to a global theatre such that massive material entities can be looked upon as if they are `contained inside this global theatre', because the deep reality is much vaster, not contained in this theatre. It consists not only of many more dimensions, but also of a much bigger complexity -- and more specifically an aspect of the bigger complexity is that each entity comes with `its personal time-space', and the different `time-spaces' of the different entities do not just nicely fit together in the one theatre of time-space --, and it is this complexity which is `rotated away' by adding spin to the individual entities. This `rotating away' makes it possible for the individual entities to smoothly enter the theatre where time-space globally reigns, and behave as if they were contained in it. Let me describe the analogous phenomenon in human cognition. We compare the theatre where time-space reigns with a well-defined space of discourse, e.g. a political agenda. For individual words or parts of sentence to `fit' in this space of discourse, as if they are its sub-elements, these individual words and parts of sentence need to carry twists, the human cognitive equivalent of spin in the physical realm, that make them appropriate and fit for this space of discourse.

In this respect, a more general way to approach what we have analyzed in the present article would be to see the duality between `time' and `space' as a duality between 'time' and `outcome set' of a measurement. If a measurement can yield different outcomes, it is the outcomes that define the realm where space can form a theatre, while the measurement itself takes place in time. In this respect, we believe to have developed an approach -- founded on the creation discovery view related to the hidden measurement approach, but generalized to any type of number of outcomes -- that can, in a more general way, `square things' such that `spin is rotated away', and time-space is defined for the measurement in consideration, except that the dimensions of the space realm will be greater than three, in general equal to $n^2-1$, where $n$ is the number of possible outcomes \cite{aertssassoli2014}.

Our aim is to develop further the new view on relativity that we have put forward here, and investigate in depth the ideas that we have presented.


\begin{thebibliography}{}

\bibitem{einstein1905} Einstein, A. (1905). Zur Elektrodynamik bewegter K\"orper. {\it Annalen der Physik}, {\bf 322}, pp. 891-921.

\bibitem{einstein1905b} Einstein, A. (1905). Ist die Tr\"agheit eines K\"orpers von seinem Energieinhalt abh\"angig? {\it Annalen der Physik}, {\bf  18}, pp. 639-642.



\bibitem{minkowski1915} Minkowski, Hermann (1915). Das Relativit\"atsprinzip. {\it Annalen der Physik}, {\bf 352}, pp. 927-938. 

\bibitem{einstein1916} Einstein, A. (1916). Die Grundlage der allgemeinen Relativit\"atstheorie. {\it Annalen der Physik}, {\bf  354}, pp. 769-822.



\bibitem{einstein1920} Einstein, A. (1920). {\it Relativity: The Special and General Theory}. (translated by Lawson, R. W. from the original published in 1916). London: Methuen \& Co Ltd.

\bibitem{einstein1952} Einstein, A. (1952). {\it The Principle of Relativity}. New York: Dover Publications.

\bibitem{misnerthornewheeler1973} Misner, C. W., Thorne, K. S. and Wheeler, J. A. (1973). {\it Gravitation}. San Francisco: Freeman and Company.

\bibitem{aerts2009} Aerts, D. (2009). Quantum particles as conceptual entities: A possible explanatory framework for quantum theory. {\it Foundations of Science}, {\bf 14}, pp. 361-411.

\bibitem{aerts2010a} Aerts, D. (2010). Interpreting quantum particles as conceptual entities. {\it International Journal of Theoretical Physics}, {\bf 49}, pp. 2950-2970.

\bibitem{aerts2010b} Aerts, D. (2010). A potentiality and conceptuality interpretation of quantum physics. {\it Philosophica}, {\bf 83}, pp. 15-52.

\bibitem{aerts2013} Aerts, D. (2013). Quantum theory and conceptuality: Matter, stories, semantics and space-time. {\it Scientiae Studia}, {\bf 11}, pp. 75-100.

\bibitem{aerts2014} Aerts, D. (2014). Quantum theory and human perception of the macro-world. {\it Frontiers in Psycholology}, 24 June, 2014. \url{http://dx.doi.org/10.3389/fpsyg.2014.00554}

\bibitem{aerts1998} Aerts, D. (1998). The entity and modern physics: the creation-discovery view of reality. In E. Castellani (Ed.), {\it Interpreting Bodies: Classical and Quantum Objects in Modern Physics} (pp. 223-257). Princeton: Princeton Unversity Press.

\bibitem{aerts1999} Aerts, D. (1999). The stuff the world is made of: physics and reality. In D. Aerts, J. Broekaert and E. Mathijs (Eds.), {\it Einstein meets Magritte: An Interdisciplinary Reflection} (pp. 129-183). Dordrecht: Springer.

\bibitem{aerts1996a} Aerts, D. (1996). Framework for possible unification of quantum and relativity theories. {\it International Journal of Theoretical Physics}, {\bf 35}, pp. 2399-2416.


\bibitem{aerts1996b} Aerts, D. (1996). Relativity theory: what is reality?. {\it Foundations of Physics}, {\bf 26}, pp. 1627-1644.
 
\bibitem{dirac1928} Dirac, P. A. M. (1928). The quantum theory of the electron. {\it Proceedings of the Royal Society A: Mathematical, Physical and Engineering Sciences}, {\bf 117}, pp. 610-624. 
 
 \bibitem{aertssassoli2014} Aerts, D and Sassoli de Bianchi, M. (2014). The extended Bloch representation of quantum mechanics and the hidden-measurement solution to the measurement problem. {\it Annals of Physics}, {\bf 351},  pp. 975--1025.


\end{thebibliography}
\end{document}